\documentclass[12pt]{amsart}
\usepackage{amssymb,latexsym}

\begin{document}
\title[Can a wormhole \ldots really be traversable?]
{Can a wormhole supported by only small amounts of exotic matter
     really be traversable?}
\author{Peter K. F. Kuhfittig}
\address{Department of Mathematics\\
Milwaukee School of Engineering\\
Milwaukee, Wisconsin 53202-3109}
\date{\today}

\begin{abstract}
Recent studies have shown that: (a) quantum effects may be sufficient to 
support a wormhole throat, and (b) the total amount of ``exotic matter" 
can be made arbitrarily small.  Unfortunately, using only small amounts 
of exotic matter may
result in a wormhole that flares out too slowly to be traversable
in a reasonable length of time.  Combined with the Ford-Roman 
constraints, the wormhole may also come close to having an event 
horizon at the throat. This Brief Report examines a model that 
overcomes these difficulties, while satisfying the usual traversability 
conditions.  This model also confirms that the total amount of 
exotic matter can indeed be made arbitrarily small.
\end{abstract}

\maketitle 

PAC number(s): 04.20.Jb, 04.20.Gz

\section{Introduction}
Wormholes may be defined as handles or tunnels linking different 
universes or widely separated regions of our own universe.  That such
wormholes may be traversable by humanoid travelers was first conjectured
by Morris and Thorne~\cite{MT88}.  To hold such a wormhole open, 
violations of certain energy conditions are unavoidable \cite{MT88,
MTY88, mV96, jF97}.  As a result, the energy density of matter may be
seen as negative by some observers.  Morris and Thorne called such 
matter ``exotic."

While all classical forms of matter obey the weak energy condition (WEC)
$T_{\alpha\beta}\mu^{\alpha}\mu^{\beta}\ge0$ for all time-like vectors 
and, by continuity, all null vectors, quantum fields can generate 
locally negative energy densities, which may be arbitrarily large 
at a given point.  In addition to the WEC, wormhole spacetimes
violate the averaged null energy condition (ANEC) \cite{MTY88, FR96},
which states that $\int T_{\alpha\beta}k^{\alpha}k^{\beta}d\lambda\ge0$,
where the integral is taken along a complete null geodesic with tangent
vector $k$ and affine parameter $\lambda$.  While quantum field theory
has generously allowed the existence of exotic matter, it also 
constrains the wormhole geometries, as a detailed analysis by 
Ford and Roman \cite{FR96} has shown.  In particular, the exotic 
matter has to be confined to a shell very much thinner than the throat.

Consider now the spherically symmetric line element
\begin{equation}\label{E:line1}
   ds^2=-e^{2\gamma(r)}dt^2+\frac{dr^2}{1-b(r)/r}+r^2(d\theta^2
      +\text{sin}^2\theta\,d\phi^2).
\end{equation}
Here $\gamma(r)$ is the \emph{redshift function} and $b(r)$ the \emph
{shape function}.  Proposals to restrict the exotic matter to an 
arbitrarily thin region are discussed by Kuhfittig~\cite{pK99, pK02}
under the condition that $b'(r)$ be close to unity near the throat.
A more general discussion of an arbitrarily small energy condition
violation is presented by Visser \emph{et al.}~\cite{mV03} in 
conjunction with the main results of global analysis in classical
general relativity.  Assuming the ANEC violation, the integral
representing the total amount of exotic matter is shown to be
arbitrarily small, provided that $e^{\gamma(r_0)}\rightarrow 0$, where
$r=r_0$ is the throat of the wormhole. The limit actually refers to 
a sequence of wormholes.  (If $e^{\gamma(r)}\rightarrow 0$ 
as $r \rightarrow r_0$ for a
specific wormhole, we would be dealing with an event horizon.)

Finally, Hochberg \emph{et al.}~\cite{dH97} in discussing their 
self-consistent wormhole solution of semiclassical gravity, present
numerical evidence suggesting that quantum effects may be sufficient
to support a wormhole throat.

The purpose of this Brief Report is to show that a wormhole supported by
only minute amounts of exotic matter may be traversable in practice,
not just in principle.  We assume that the Ford-Roman constraints 
are satisfied, while avoiding an event horizon at the throat.  The
wormhole is small enough to be traversed in a reasonable length of
time;  the radial tidal forces and the gradient of the redshift
function are small enough to accommodate a humanoid traveler.  That
only small amounts of exotic matter are needed is confirmed in 
Sec.~\ref{S:traversability}.

\section{The ANEC violation}\label{S:ANEC}
In place of the line element (\ref{E:line1}) we will use the form
\begin{equation}\label{E:line2}
   ds^2=-e^{2\gamma(r)}dt^2+e^{2\alpha(r)}dr^2+r^2(d\theta^2
       +\text{sin}^2\theta\,d\phi^2).
\end{equation}
To save space we will omit the usual discussion of the basic wormhole 
features except to note that  the graph of the function 
$\alpha=\alpha(r)$ has a vertical asymptote at $r=r_0$: 
$\lim_{r \to r_0+}\alpha(r)=+\infty$.  (For further details see
Refs. \cite{MT88, pK02}.)  From the line elements above we have
\begin{equation}\label{E:shape}
       b(r)=r\left(1-e^{-2\alpha(r)}\right).
\end{equation}
One of the general bounds for wormhole geometries discussed in Ref.
~\cite{FR96}, Sec.~ V, is shown to be weakest when $b'(r)$ is close to 
unity near the throat.  Accordingly, we assume that the graph of 
$\alpha=\alpha(r)$ is steep enough near $r=r_0$ to meet this condition. 
If $b'(r)$ is close to unity near the
throat, then the embedding diagram will flare out very slowly.  This
slow flaring out need not be fatal, however, as shown in Ref.~\cite
{pK02}. (We will return to this point in the next section.)

Also from Ref. ~\cite {pK02} the WEC violation in terms of 
$\alpha$ and $\gamma$ is given by $\rho-\tau<0$, where
\begin{equation}\label{E:exotic}
  \rho-\tau=\frac{1}{8\pi}\left[\frac{2}{r}e^{-2\alpha(r)}
     \left(\alpha'(r)+\gamma'(r)\right)\right].
\end{equation}
Following Visser \emph{et al.}\,\cite{mV03}, a natural way to measure
the mass of the wormhole (including both asymptotic regions) is
\begin{equation}\label{E:volume}
   \int\nolimits_{\text{Vol}}\rho(r)dV=2\int\nolimits_{r_0}^{\infty}
      4\pi r^2\rho(r)dr.
\end{equation}
Because of the ANEC violation, our interest centers mainly on
the integral $\int\nolimits_{\text{Vol}}(\rho-\tau)dV$.
Since Eq. (\ref{E:exotic}) can be written
\[
  \rho-\tau=\frac{1}{8\pi r^2}\left[2re^{-2\alpha(r)}
      (\alpha'(r)+\gamma'(r))\right],
\]
it follows from Eq. (\ref{E:volume}) that 
\begin{equation}
   \int\nolimits_{\text{Vol}}(\rho-\tau)dV=2\int\nolimits_{r_0}
       ^{\infty}r e^{-2\alpha(r)}
          (\alpha'(r)+\gamma'(r))dr.
\end{equation}
Integrating by parts,
\begin{equation}\label{E:ANEC1}
   \int\nolimits_{\text{Vol}}(\rho-\tau)dV=-2\int\nolimits_{r_0}
     ^{\infty}(\alpha(r)+\gamma(r))e^{-2\alpha(r)}(1-2r\alpha'(r))dr
\end{equation}
since the boundary term vanishes at the throat due to the factor 
$e^{-2\alpha(r)}$ and at infinity due to the asymptotic behavior.

A good choice for $\alpha(r)$ is
\[
   \alpha(r)=\frac{K^n}{(r-r_0)^n},\quad n\ge1,
\]
for some constant $K$ having the same units as $r$.  The condition
$n\ge 1$ ensures that $b'(r)\approx 1$ near the throat.  To avoid an 
event horizon, we let
\[
    \gamma(r)=-\frac{L^n}{(r-r_2)^n},\quad n\ge1,
\]
for some constant $L$ and where $r_2$ is such that $0<r_2<r_0$.

To satisfy the Ford-Roman constraints, the WEC must be satisfied 
outside the interval $[r_0,r_1]$ for some $r_1$.  To accomplish this, 
construct $\alpha$ and $\gamma$ so that $\left|\alpha'(r_1)\right|=
\left|\gamma'(r_1)\right|$.  Now let $\gamma_1(r)=-\gamma(r)$, choose
a suitable $K$, and determine $L$ so that $\alpha'(r)=\gamma_1'(r)$.
The result is
\begin{equation}\label{E:Ford}
   L^n=\left[\frac{(r_1-r_2)^{n+1}}{(r_1-r_0)^{n+1}}\right]K^n.
\end{equation} 
With this choice of $L$ it is easy to show that $\left|\alpha'(r)\right|    
>\left|\gamma'(r)\right|$ for $r_0<r<r_1$;  more precisely,
\[
   \alpha'(r)=-\frac{nK^n}{(r-r_0)^{n+1}}<-\frac{n(r_1-r_2)^{n+1}K^n}
     {(r_1-r_0)^{n+1}}\frac{1}{(r-r_2)^{n+1}}=\gamma'(r).
\]
To the right of $r=r_1$ the inequality is reversed.
As a result, we have $\rho-\tau<0$ in the interval $(r_0,r_1)$ and
$\rho-\tau\ge0$ for $r\ge r_1$.  Also, since the exotic matter is 
confined to the spherical shell extending from $r=r_0$ to $r=r_1$, we
have
\begin{equation}\label{E:ANEC2}
    \int\nolimits_{\text{Shell}}(\rho-\tau)dV<0,
\end{equation}
which represents the ``total amount" of energy-condition
violating matter.
One of our goals is to show that the integral (\ref{E:ANEC2}) can be 
made arbitrarily small.  To that end observe that by the mean-value
theorem there exists a number $c\in(r_0,r_1)$ such that
\begin{multline}\label{E:ANEC3}
   \int\nolimits_{\text{Shell}}(\rho-\tau)dV=-2\int\nolimits_{r_0}
   ^{r_1}\left(\alpha(r)+\gamma(r)\right)e^{-2\alpha(r)}
             \left(1-2r\alpha'(r)\right)dr\\
   =-\frac{2}{r_1-r_0}(\alpha(c)+\gamma(c))e^{-2\alpha(c)}
         (1-2c\alpha'(c))\\
    =-\frac{2}{r_1-r_0}\left(\frac{K^n}{(c-r_0)^n}-\frac{L^n}
         {(c-r_2)^n}\right)e^{-2K^n/(c-r_0)^n}\\
            \times\left(1+2cnK^n\frac{1}{(c-r_0)^{n+1}}\right).
\end{multline}
To see why this integral may be vanishingly small, let $r_1\rightarrow
r_0$ and hence $c\rightarrow r_0$.  By l'Hospital's rule the right 
side of Eq.~ (\ref{E:ANEC3}) approaches $0$. Unfortunately, since the
construction of $\gamma(r)$ depends on $r_1$, this limit cannot be taken
directly.  In fact, for a fixed $\alpha(r)$, if $r_1\rightarrow r_0$,
$\gamma'(r_1)$ gets ever larger, causing
the sequence $\gamma(r_0)$ would recede to $-\infty$, so that 
$e^{\gamma(r_0)}\rightarrow 0$, creating the very event horizon that
we are trying to avoid.

We will return to this problem at the end of the next section.

\section{Traversability conditions}\label{S:traversability}
As noted earlier, $b'(r)$ is close to unity near the throat.  The 
resulting slow flaring out could make the wormhole too large to be
traversable in a reasonable length of time.  To analyze this problem,
as well as the lateral tidal constraint and the gradient of the 
redshift function, we assume the following:  with Eq. (\ref{E:Ford})
in mind, suppose for now that $r_2$ has been chosen so that $L$ is not
much larger than $K$ as long as $n$ is small.  That way we can use the
same values for $L$ and $K$ and consider the case $L\ne K$ later.  Also,
since the wormholes are likely to be very large compared to $r_0$, we
may assume that $r_0$, $r_1$, and $r_2$ are negligible for the purpose
of estimating larger distances.  As a result, $\gamma(r)=-\alpha(r)$;  
observe that the Ford-Roman constraints are trivially satisfied.

For our first model, we choose $n=1$, so that $\alpha(r)=K/r$ and
$\gamma(r)=-L/r=-K/r$.  For the traversability conditions we follow 
Morris and Thorne \cite{MT88}.  The crucial radial tidal constraint 
is given by
$\left|R_{\hat{1}'\hat{0}'\hat{1}'\hat{0}'}\right|
        =\left|R_{\hat{r}\hat{t}\hat{r}\hat{t}}\right|\le 
        (10^8 \,\text{m})^{-2}$.
By direct calculation or from Ref. \cite{pK02}
\begin{multline}\label{E:tidal}
   \left|R_{\hat{r}\hat{t}\hat{r}\hat{t}}\right|
    =\left|e^{-2\alpha(r)}\left[\gamma''(r)-\alpha'(r)\gamma'(r)
     +(\gamma'(r))^2\right]\right|\\
       =\left|e^{-2K/r}\left(-\frac{2K}{r^3}+\frac{2K^2}{r^4}
             \right)\right|.
\end{multline}           
The function on the right (inside the absolute value signs)
attains respective minimum and maximum values at $r=\frac{1}{3}(3\pm
\sqrt{3})K$.  To meet the constraint at these values, we choose 
$K=5.0\times 10^{-9}$ l.y. (light year).

Concerning the size of the wormhole as measured by the placement
of the space stations, if 
we choose $r=0.00006964$ l.y.\,$\approx 6.6\times 10^8$ km, then
\[
   \gamma'(r)=\frac{K}{r^2}=\frac{5.0\times 10^{-9}\times 9.46\times
     10^{15}\,\text{m}}{(0.00006964\times 9.46\times 10^{15}
       \,\text{m})^2}\,\approx 1.1\times 10^{-16}\,\text{m}^{-1},
\]
which meets the constraint 
$\left|\gamma'(r)\right|\le g_{\oplus}/(c^2\sqrt{1-b(r)/r})$.
Finally, $b(r)/r$ is well within 1\% of unity, also recommended 
in Ref. \cite{MT88}.

The distance $r=6.6\times10^8$ km\,$\approx$ 4 A.U. may seem rather large, but if we
assume, as suggested in Ref.~\cite{MT88}, that the spaceship 
accelerates at $g_{\oplus}=9.8\,\text{m}/\text{s}^2$ halfway to the
throat and decelerates at the same rate until it comes to rest at the
throat, then the throat would be reached in only about 6 days.

It is instructive to compare this model to one for which $n=2$:  let
$\alpha(r)=K^2/r^2$ and $K=1.0\times 10^{-8}$ l.y.  All the above conditions
are met, but the size of the wormhole is now only $0.000005789$ l.y.
\,$\approx\,\frac{1}{3}$ A.U.

For completeness let us momentarily suspend the condition that $b'(r)$
gets close to unity as we approach
the throat: assume that $b(r)=r_0=2m$ outside a thin region
extending from the throat to $r=a$, as in Ref. \cite{mV03}; we further 
assume that $a<r_1$.  Then
\[
  \alpha(r)=-\frac{1}{2}\,\text{ln}(1-\frac{2m}{r});
\]
for the redshift function let $\gamma(r)=-K/r$.  (Both $K$ and $m$ are
measured in meters.)  As in the previous section, we need to find
$r=r_1$ such that
\[
   \alpha'(r_1)+\gamma'(r_1)=-\frac{m}{(1-2m/r_1)
    r_1^2}+\frac{K}{r_1^2}=0,
\]
whence 
\[
       K=\frac{m}{1-2m/r_1}.
\]
If $r_{\text{th}}$ is the thickness of the shell, then $r_{\text{th}}
=r_1-r_0=r_1-2m$, and
\[
  K=\frac{m}{1-2m/(2m+r_{\text{th}})}=m\left(2m\,r_{\text{th}}^{-1}
    +1\right).
\]
The factor $r_{\text{th}}^{-1}$ will cause $K$ to be large for any
reasonable value of $m$.  For example, Ref.~\cite{MTY88}, which
discusses a wormhole based on the experimentally confirmed Casimir
effect, gives $r_{\text{th}}=10^{-12}$ m.  As a consequence, the radial 
tidal constraint, Eq. (\ref{E:tidal}), does not even come close to being 
satisfied at any point not too far from the throat.  The primary reason
is that the coefficient $e^{-2\alpha(r)}$ collapses to $1-2m/r$, which
does not decay fast enough as $r\rightarrow 2m$.  The resulting wormhole
is therefore not traversable by humanoid travelers.  (It is not hard to 
show that this conclusion holds for any differentiable function
$\gamma =\gamma(r)$, particularly near $r=r_1$.)

\vspace{16pt}
In obtaining the estimates in this section, we made the 
simplifying assumption that $L=K$.  An
obvious alternative is to choose a smaller $K$ to start with:  referring 
to Eq. (\ref{E:Ford}), if we let 
\[
   A(n)=\frac{(r_1-r_2)^{n+1}}{(r_1-r_0)^{n+1}},
\]
then $L^n=A(n)K^n$.  (The expression for $A(n)$ shows that $r_2$ 
should be restricted so that $A(n)$ does not become excessively 
large.)  Replacing $K^n$ by $K^n_{\text{new}}=K^n/A(n)$ 
satisfies the requirements of Eq.~ (\ref{E:Ford}) while
keeping $L^n$ and $\gamma(r)$ intact.  As long as $n$ is small, 
changing $\alpha$ to $\alpha(r)=
K^n_{\text{new}}/r^n$ will not have a drastic effect on the above 
estimates, as can be seen from Eq.~ (\ref{E:tidal}).  

For large $n$ these estimates, as well as the resulting model, become 
clearly invalid:  since $A(n)$ keeps increasing for any fixed 
$r_2$, $K_{\text{new}}^n=K^n/A(n)$ is a decreasing sequence.  
The exponential function $e^{-2\alpha(r)}$ will therefore begin 
to decay too slowly for the radial tidal constraint to be satisfied.  
(If the model were valid for large $n$,
then the size of the wormhole could be decreased indefinitely,
which would not make physical sense.)

Returning next to Eq.~ (\ref{E:ANEC3}), if we model the wormhole by 
using the functions in this section, we can estimate the size of
$\int\nolimits_{\text{Shell}}(\rho-\tau)dV$
directly.  Observe that, since $c<r_1$,
\begin{multline}\label{E:ANEC4}
   \left|\int\nolimits_{\text{Shell}}(\rho-\tau)dV\right|
   =\left|-\frac{2}{r_1-r_0}(\alpha(c)+\gamma(c))e^{-2\alpha(c)}
       (1-2c\alpha'(c))\right|\\
    < \left|-\frac{2}{c-r_0}\left(\frac{K^n}{(c-r_0)^n}-
       \frac{L^n}{(c-r_2)^n}\right)e^{-2K^n/(c-r_0)^n}\right.\\
        \left.\times\left(1+2cnK^n\frac{1}{(c-r_0)^{n+1}}\right)\right|.
\end{multline}       
Simple calculator trials show that for any reasonable choice of $r_1$ 
(and hence of c),
\[
    \left|\int\nolimits_{\text{Shell}}(\rho-\tau)dV\right|\ll 10^{-100},
\]   
even for $n=1$.  In fact, this extreme inequality holds even if $L^n$
is much larger than $K^n$ and $r_1$ many orders of magnitude larger 
than the value allowed by the Ford-Roman constraints.

While the conclusion seems to depend on our choices of $\alpha$ and
$\gamma$, it is unlikely that any other acceptable choices  
would alter the results
significantly, primarily because the basic features remain the same:
both functions are assumed to be twice differentiable and hence 
continuous in their
respective domains, $\gamma(r_0)$ is finite, and $\lim_{r\to r_0+}
\alpha(r)=+\infty$.  So by continuity, 
$\lim_{r_1\to r_0}\gamma(r_1)e^{-2\alpha(r_1)}=0$,
even if $\gamma(r)$ assumes a completely different form from the one
considered earlier, while
\[
   \lim_{r_1\to r_0}\alpha(r_1)e^{-2\alpha(r_1)}=
       \lim_{r_1\to r_0}\frac{\alpha(r_1)}{e^{2\alpha(r_1)}}=0
\]
by l'Hospital's rule, suggesting that the relevant quantities are
indeed vanishingly small.

\section{Conclusion}
It is shown in this Brief Report that a wormhole held open with only small
amounts of exotic matter may be traversable, not just in principle, but
also in practice:  in the model discussed the wormhole size permits 
traversal in a reasonable length of time, while satisfying the usual
traversability conditions.  The model also accommodates the 
Ford-Roman constraints---without introducing an event horizon at the
throat.  The integral measuring the total amount of exotic matter 
proved to be vanishingly small, confirming the results in Ref.~\cite
{mV03}.

\end{document}